\documentclass[twocolumn,aps,prc,showpacs,superscriptaddress,preprintnumbers,floatfix,nofootinbib]{revtex4}
\usepackage{epsfig,graphics}
\usepackage{graphicx}% Include figure files
\usepackage{subfigure}
\usepackage{dcolumn}% Align table columns on decimal point
\usepackage{bm}% bold math
\usepackage{amsmath}
\usepackage[usenames]{color}
\usepackage{ulem} %% for strike-through
\usepackage{fancyhdr}
\usepackage{hyperref}

\newcommand{ \mt } {$m_T $}

\newcommand{ \gth } {$> $}

\begin{document}
%\begin{CJK*}{GBK}{song}

%\renewcommand {\baselinestretch 2}

\title{ Nuclear modification factor in intermediate-energy heavy-ion collisions }
\author{M. Lv}
\affiliation{Shanghai Institute of Applied Physics, Chinese
Academy of Sciences, Shanghai 201800, China}
\affiliation{University of Chinese Academy of Sciences, Beijing 100049, China}
\author{Y. G. Ma}
\thanks{Author to whom all correspondence should be addressed. Email: ygma@sinap.ac.cn}
\affiliation{Shanghai Institute of Applied Physics, Chinese
Academy of Sciences, Shanghai 201800, China}\textcolor[rgb]{1.00,0.00,0.00}{}
\author{G. Q. Zhang}
\affiliation{Shanghai Institute of Applied Physics, Chinese
Academy of Sciences, Shanghai 201800, China}
\author{J. H. Chen}
\affiliation{Shanghai Institute of Applied Physics, Chinese
Academy of Sciences, Shanghai 201800, China}
\author{D. Q. Fang}
\affiliation{Shanghai Institute of Applied Physics, Chinese
Academy of Sciences, Shanghai 201800, China}

\begin{abstract}
The transverse momentum dependent nuclear modification factors (NMF), namely $R_{CP}$, is investigated for protons produced in Au + Au at 1$A$ GeV within the framework of the isospin-dependent quantum molecular dynamics (IQMD) model. It is found that the radial collective motion during the expansion stage affects the NMF at low transverse momentum a lot. By fitting the transverse mass spectra of protons with the distribution function from the Blast-Wave model, the magnitude of radial flow can be extracted. After removing the contribution from radial flow, the $R_{CP}$  can be regarded as a thermal one and is found to keep unitary at transverse momentum lower than 0.6 GeV/c and enhance at higher transverse momentum, which can be attributed to Cronin effect.
\end{abstract}
\pacs{24.10.-i, 25.75.Ld}

\date{ \today}
\maketitle
%%%%%%%%%%%%%%%%%%%%%%%%%%
%Outline
%  1 Introduction
%  2 IQMD
%  3 mT spectra
%    3.1 boltzmann fit
%    3.2 blast wave fit
%  4 nuclear modification factor
%    4.1 the normal Rcp
%    4.2 the thermal Rcp
%  5 conclusion
%%%%%%%%%%%%%%%%%%%%%%%%%%

%%%%%%%%%%%%%%%%%%%%%%%%%%%%%%%%%%%%%%%%%%%%%%%%%%%%%%%%%%%%%%%%%%%%%%%%%%%%%%%%%%%%%%%%%%%%%%%%%%%%%%%%%%%%%%%%%%%%%%%%%%%%%%%%%%%%%%%%%
%%%%%%%%%%%%%%%%%%%%%%%%%%%%%%%%%%%%%%%%%%%%%%%%%%%%%%%%%%%%%%%%%%%%%%%%%%%%%%%%%%%%%%%%%%%%%%%%%%%%%%%%%%%%%%%%%%%%%%%%%%%%%%%%%%%%%%%%%
\section{Introduction}

Recently, the nuclear modification factor (NMF) has been extensively investigated  for  different particles at various collision energies in relativistic heavy-ion collisions (HIC) ~\cite{RHIC_NMF1,RHIC_NMF2,RHIC_NMF3,RHIC_NMF4}. These studies indicate that the NMF, which can be represented either by the modification factor between nucleus-nucleus (AA) collisions and proton-proton (pp) collision $R_{AA}$ or the one between the central collisions and peripheral collisions $R_{CP}$, is very useful for the study of the quantitative properties of the nuclear medium response when the high speed jet transverses it. In high transverse momentum ($p_{T}$) region, NMF is suppressed owing to jet quenching effect in hot-dense matter and thus has become one of the robust evidences on the existence of the Quark-Gluon-Plasma \cite{WhitePaper,Jet2}. In lower $p_{T}$ region, radial flow boosts or the Cronin Effect~\cite{CroninE1} competes with the quenching effect and enhances the NMF, which has been also demonstrated  by the Relativistic Heavy-Ion Collider (RHIC) beam energy scan (BES) project~\cite{Horvat_BES}.

Meanwhile, collective motion plays an important role in the time evolution of particles, which has been studied over a wide range of collision energy in heavy-ion collision. Around 1$A$ GeV incident energy in central HIC, the colliding nuclei are expected to be stopped and lead to densities of $2\sim3\rho_0$ ($\rho_0$ is the normal nuclei density) at the largest compression time~\cite{HICs}. At this point, high excitation energy stage is reached and some parts of the excitation energy are converted into collective motion, such as radial flow ~\cite{Siemens,flow_evi2}. Thus, in the following expansion stage, the products move outward containing both the collective motion and the thermal motion. It will be of highly interesting to decouple these two parts, because each of them reflects important information of the HIC process. Efforts have been made by the Blast-Wave model~\cite{Lisa_rflow,FOPI_rflow1,FOPI_rflow2}, and the collective motion parameter (radial flow velocity) together with the thermal motion parameter (temperature) can be extracted at the same time.

Nevertheless, till now, NMF has not been investigated in intermediate energy HIC yet to our knowledge. In particular, the $R_{CP}$ shape will be strongly affected by the radial flow which plays a significant role in the expansion stage especially for the low $p_{T}$ particles. In order to understand the properties of the nuclear medium response, one may want to know what the behavior of $R_{CP}$ without the contribution from radial flow might be. In the present paper, we address NMF in intermediate energy HIC to study the properties of nuclear medium for the first time. After removing the contribution from the radial flow, the NMF can be regarded as a thermal one, which reflects property of thermal medium  produced in intermediate energy HIC. The $R_{CP}$ for emitted protons in Au + Au collision at 1$A$ GeV is investigated systematically.

The article is organized as follows: A brief introduction about the IQMD model is given in Sec. II. In Sec.III, we compare the kinetic energy spectra of light fragments obtained by the IQMD with the one from the EOS experimental result, then we fit the proton transverse mass (\mt) spectra with the Boltzmann function and with the distribution function from Blast-Wave model. In Sec. IV, we show the results of NMFs and compare them with the results extracted from the KaoS experimental data. Finally, the thermal $R_{CP}$ is recalculated by removing the contribution from radial flow. Summary and conclusion are presented in Sec. V.

%%%%%%%%%%%%%%%%%%%%%%%%%%%%%%%%%%%%%%%%%%%%%%%%%%%%%%%%%%%%%%%%%%%%%%%%%%%%%%%%%%%%%%%%%%%%%%%%%%%%%%%%%%%%%%%%%%%%%%%%%%%%%%%%%%%%%%%%%
%%%%%%%%%%%%%%%%%%%%%%%%%%%%%%%%%%%%%%%%%%%%%%%%%%%%%%%%%%%%%%%%%%%%%%%%%%%%%%%%%%%%%%%%%%%%%%%%%%%%%%%%%%%%%%%%%%%%%%%%%%%%%%%%%%%%%%%%%
\section{Brief description of IQMD model}

The Quantum Molecular Dynamics model is a transport model which is based on a many body theory to describe  heavy ion collisions from low (dozens of MeV) to relativistic energy~\cite{Archelin_QMD,Archelin_MDQMD,Hartnack_QMD,Hartnack_IQMD}. The Isospin dependent Quantum Molecular model (IQMD), was extended from the QMD model, with considering the isospin effects. In the past decades, many applications  have been successfully performed into nuclear physics studies with the help of IQMD. For instance, IQMD has been successfully applied  to treat collective flow, multi-fragmentation, isospin effects in HIC, transport coefficient in HIC, giant resonance, and strangeness production etc \cite{Rev1,Zhou,Kumar,NST,NST2}.
%[Hartnack_EPJA_IQMD]
%\subsection{IQMD model}
In the IQMD model, the wave function of each nucleon is described as a coherent state with the form of Gaussian wave packet,
       \begin{equation}
       \phi_i(\vec{r},t)= \frac{1}{(2\pi L)^{3/4}}exp(-\frac{(\vec r-\vec r_i(t))^{2}}{4L})exp(\frac{i\vec r\cdot \vec p_i(t)}{\hbar}).
       \label{eq1}
       \end{equation}
Here $r_i$ and $p_i$ are the time dependent variables which describe the center of the packet in coordinate and momentum space, respectively. The parameter $L$, related to the width of  wave packet in coordinate space, is determined by the size of reaction system, i.e. $L$ = 1.08 fm$^{2}$ for Ca+Ca system and $L$ = 2.16 fm$^{2}$ for Au+Au system in this work. The wave function of the system is the direct product of all the nucleon wave functions without considering the Fermion property of nucleon:
     \begin{equation}
      \Phi(\vec{r},t) = \prod_i \phi_i (\vec{r},t).
      \label{eq1.5}
      \end{equation}
As a compensation, Pauli blocking is employed in the initializations and collision process to restore some parts of the quantum property of many Fermion system.

By applying a generalized variational principle on the action of the many-body system, one can get the  equations of motion for $p_i$ and $r_i$, which are listed as follows
       \begin{equation}
       \vec p_i=-\frac{\partial \left\langle H \right\rangle}{\partial \vec r_i};\\
       \vec r_i=\frac{\partial \left\langle H \right\rangle}{\partial \vec p_i}.
       \label{eq2}
       \end{equation}

The Hamiltonian $\left\langle H \right\rangle=\left\langle T \right\rangle+\left\langle V \right\rangle$ where $T$ is the kinetic energy, the potential $V$ is expressed by
\begin{equation}
\label{meanfield}
\langle V \rangle = \frac{1}{2} \sum_{i} \sum_{j \neq i}
 \int f_i(\vec{r},\vec{p},t) \,
V^{ij}  f_j(\vec{r}\,',\vec{p}\,',t)\, d\vec{r}\, d\vec{r}\,'
d\vec{p}\, d\vec{p}\,'.
\end{equation}
In the above, the Wigner distribution function $f_i
(\vec{r},\vec{p},t)$, which is the phase-space density of the $i$th
nucleon, is obtained by applying the Wigner transformation on the
single nucleon wave function:
       \begin{equation}
       f_i(\vec r,\vec p,t)=\frac{1}{(\pi\hbar)^{3}}e^{-(\vec r-\vec r_i(t))^{2}\frac{1}{2L}} e^{-(\vec p-\vec p_i(t))^2\frac{2L}{(\hbar)^{2}}}.
       \label{eq3}
       \end{equation}

The baryon-potential consists of the real part of the G-Matrix which is supplemented by the Coulomb interaction between the charged particles. The former one can be divided into three parts, the Skyrme-type interaction, the finite-range Yukawa potential, and the momentum-dependent interaction (MDI) parts.
The two-body interaction potential $V^{ij}$ in Eq.~\ref{meanfield} can be expressed as follows:
       \begin{equation}
       \begin{split}
       V^{ij}&=G^{ij}+V_{Coul}^{ij}=V_{Skyme}^{ij}+V_{Yuk}^{ij}+V_{MDI}^{ij}+V_{Coul}^{ij}\\
       &=t_{1}\delta(\vec x_{i}-\vec x_{j})+t_{2}\delta(\vec x_{i}-\vec x_{j})\rho^{\gamma-1}(\vec x_{i})\\
       &+t_{3}\frac{exp(-(\vec x_{i}-\vec x_{j})/\mu)}{(\vec x_{i}-\vec x_{j})/\mu}  \\
       &+t_{4}ln^{2}[1+t_{5}(\vec p_{i}-\vec p_{j})^{2}]\delta(\vec x_{i}-\vec x_{j}) \\
       &+\frac{Z_{i}Z_{j}e^{2}}{\vec x_{i}-\vec x_{j}}.
       \label{eq4}
       \end{split}
       \end{equation}
The symmetry potential between protons and neutrons corresponding to the Bethe-Weizsacker mass formula can be taken as
       \begin{equation}
       V_{sym}^{ij}=t_6\frac{1}{\rho_0}T_{3i} T_{3j} \delta(r_i-r_j)
       \label{eq5}
       \end{equation}
with $t_6$=100 MeV. By integrating Skyrme part as well as the momentum dependent part of the two-body interaction and introducing the interaction density,
       \begin{equation}
       \rho_{ij}=\frac{1}{(4\pi L)^{3/2}} \sum_{j\neq i} exp[-\frac{(\vec r_{i}-\vec r_{j})^2}{4L}],
         \label{eq6}
       \end{equation}
one can get the local mean field potential which contains the Skyrme potential and momentum dependent potential
       \begin{equation}
       U=\alpha(\frac{\rho}{\rho_0})+\beta(\frac{\rho}{\rho_0})^{\gamma}+\frac{\rho}{\rho_0} \int d\vec{p}\,'g(\vec{p}\,') \delta\ln^2[\epsilon(\vec{p}-\vec{p}\,')^2+1],
       \label{eq7}
       \end{equation}
where $\rho_0$ is the saturation density at ground state, $g(\vec p,t)=\frac{1}{(\pi\hbar)^{3/2}}\sum_{i}e^{-(\vec p-\vec p_i(t))^2\frac{2L}{(\hbar)^{2}}}$ is the momentum distribution function, the interaction density $\rho=\sum_{ij}\rho_{ij}$, and $\alpha$, $\beta$, and $\gamma$ are the Skyrme parameters, which connect tightly with the EOS of the bulk nuclear matter, as listed in Table~\ref{qmd_par}.

       \begin{table}[htbp]
       \caption{Parameter sets for the nuclear equation of state used in the IQMD model. S and H represent the soft and hard equation of state, respectively, M refers to the inclusion of momentum dependent interaction. This table is adapted  from~\cite{Hartnack_IQMD}.}
       \label{qmd_par}
       \centering
       \begin{tabular}{p{35pt} p{38pt} p{38pt} p{38pt} p{38pt} p{38pt}}
          \hline
          \hline
             & $\alpha$ & $\beta$ & $\gamma$ & $\delta$ & $\epsilon$ \\
             & (MeV)    & (MeV)   &          &  (MeV)   & $(\frac{c^{2}}{(GeV)^{2}})$ \\
          \hline
          S  & -356 & 303 & 1.17 & -    & -   \\
          SM & -319 & 320 & 1.14 & 1.57 & 500 \\
          H  & -124 & 71  & 2.00 & -    & -   \\
          HM & -130 & 59  & 2.09 & 1.57 & 500 \\
          \hline
          \hline
       \end{tabular}
       \end{table}

With the help of coalescence mechanism, the information of fragments produced in HICs can be identified in IQMD. A simple coalescence rule to form a fragment is used with the criteria $\Delta r$ = 3.5 fm and $\Delta p$ = 300 MeV/c between two considered nucleons.

%%%%%%%%%%%%%%%%%%%%%%%%%%%%%%%%%%%%%%%%%%%%%%%%%%%%%%%%%%%%%%%%%%%%%%%%%%%%%%%%%%%%%%%%%%%%%%%%%%%%%%%%%%%%%%%%%%%%%%%%%%%%%%%%%%%%%%%%%
%%%%%%%%%%%%%%%%%%%%%%%%%%%%%%%%%%%%%%%%%%%%%%%%%%%%%%%%%%%%%%%%%%%%%%%%%%%%%%%%%%%%%%%%%%%%%%%%%%%%%%%%%%%%%%%%%%%%%%%%%%%%%%%%%%%%%%%%%
\section{transverse mass spectra}

\begin{figure}[htbp]
       \includegraphics[width=0.48\textwidth]{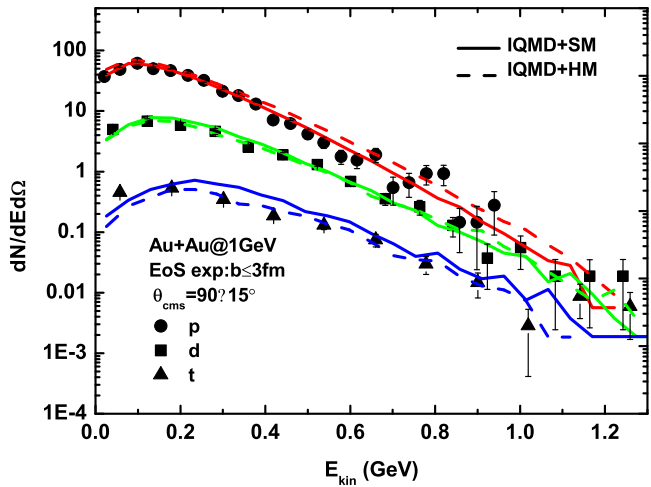}\\
       \caption{(Color online)  $E_{kin}$ spectra of proton (circle), deuteron (square), triton (triangle) in Au + Au at 1$A$ GeV. The solid symbols  are the experimental data from the EOS collaboration~\cite{Lisa_rflow}, and the solid line is  our IQMD simulation with the soft+MDI potential, and the dash line is for the hard+MDI potential.}
       \label{pr_Ekin}
       \end{figure}

The Au+Au collisions at 1$A$ GeV are simulated with the IQMD model for both the soft equation of state with MDI (SM) and the  hard equation of state with MDI (HM).
In order to test the reliability of the model, the kinetic spectra of proton, deuteron  and triton obtained by the IQMD with the above equation of state situations are compared with the EOS experimental data in Fig~\ref{pr_Ekin}~\cite{Lisa_rflow}. The conditions of the chosen fragments in our model calculations are $b \leq 3 fm, \theta_{cms}=90\pm 15^{\circ}$, which are the same as the EOS experimental situation. As shown in Fig~\ref{pr_Ekin}, the solid lines are  the spectra extracted from the IQMD with soft and MDI potential, and the dash lines are  the spectra extracted with HM potential. From Fig.~\ref{pr_Ekin}, we find that both the soft and hard equation of state situation can essentially describe the EOS experimental data nicely. A little higher yield of protons and lower yield of  t are observed in the case of the HM potential. The enhanced yield of protons can be attributed to  stronger mean field interaction in HM case compared to that in SM case.

\begin{figure}
       \includegraphics[width=0.50\textwidth]{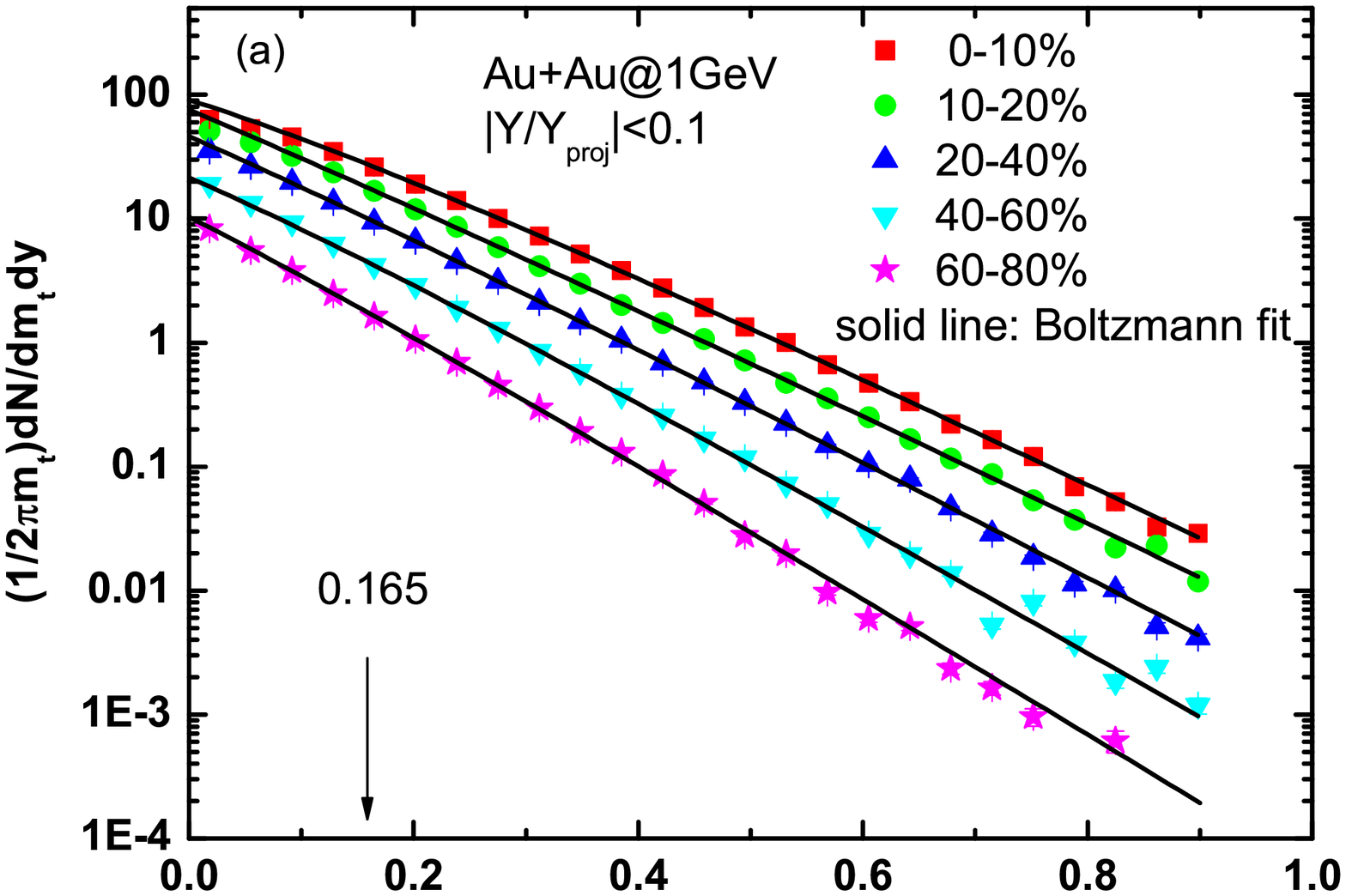}\\
       \includegraphics[width=0.50\textwidth]{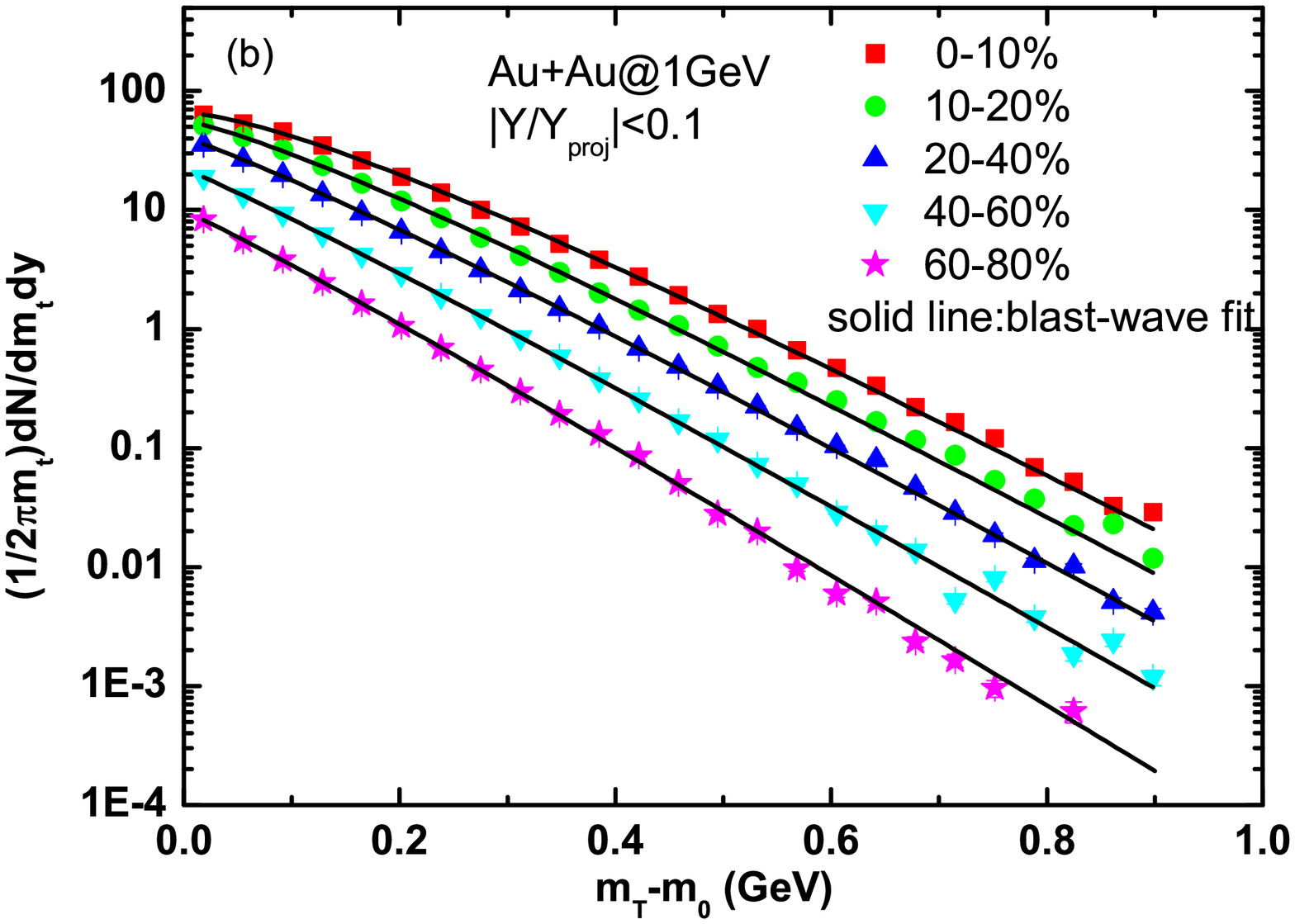}\\
       \caption{(Color online)  $m_{T}$ spectra of protons at different centralities (0-10\%,10-20\%,20-40\%,40-60\%,60-80\%) for Au + Au at 1$A$ GeV from IQMD+SM simulation, the upper panel is the Boltzmann fit in the range of $m_{T} >$ 0.165 GeV , and the lower panel is the Blast-Wave fit.}
       \label{pr_fit}
       \end{figure}

Transverse mass $m_{T}$, is given by $m_{T}=\sqrt{p_{T}^2+m_{0}^2}$, with the rest mass of protons $m_{0}$. Fig.~\ref{pr_fit} shows our simulation results for the double differential transverse mass spectra ($\frac{d^{2}N}{2\pi m_{T}dm_{T}dy}$) of protons at different centralities ($0-10\%, 10-20\%,20-40\%, 40-60\%, 60-80\%$) together with two fit results either from Boltzmann distribution function or from Blast-Wave model distribution function in the framework of IQMD model with SM potential.
%The soft equation of state with MDI (SM) has been chosen owing to little difference between %soft and hard EOS as shown in Fig~\ref{pr_Ekin}.
The c.m.s. rapidity cut is $|Y/Y_{proj}|<0.1$ where $Y_{proj}$ is the initial projectile rapidity. At the end of this section, a discussion will be given for the fit parameters, i.e. temperature and radial flow.\\

%%%%%%%%%%%%%%%%%%%%%%%%%%%%%%%%%%%%%%%%%%%%%%%%%%%%%%%%%%%%%%%%%%%%%%%%%%%%%%%%%%%%%%%%%%%%%%%%%%%%%%%%%%%%%%%%%%%%%%%%%%%%%%%%%%%%%%%%%
\subsection{The Boltzmann fit}

In heavy-ion collisions, particles collide with each other randomly, which can be described in term of thermal motion ~\cite{Fermi}. Lots of works demonstrate that the Boltzmann distribution can roughly describe the particle spectra by a thermal source in HICs~\cite{thermometry,CMuntz,slopeT2,slopeT3}. If the particles are ejected from a single pure high temperature thermal source, its transverse mass spectra should satisfy the Boltzmann distribution function, i.e. $dN/(dm_{T}dy) \approx f(m_{T})\cdot exp(-m_{T}/T)$, which is close to a straight line when the y-axis is plotted logarithmically. The results of the $m_{T}$ spectra of protons fitted with the Boltzmann function are shown in Fig.~\ref{pr_fit}(a). The proton spectra at different centralities are fitted by the Boltzmann function in the region of $m_{T}$ \gth 0.165 GeV and extrapolated to the low $m_{T}$ region. It is interesting to find that the spectra in central collisions have an obvious ``shoulder"-like structure in the low $m_{T}$ region while  there is no such structure for peripheral collisions. This means the spectra in central collisions deviate from the Boltzmann distribution in the low $m_{T}$ region, which can be essentially attributed to the radial flow in central collisions. The fit parameters of the slope temperature are listed in the second column of Table ~\ref{bzfit_par}. From peripheral to central collision the slope temperature extracted from the Boltzmann function fit shows a decreasing trend, which is very different from that extracted from the Blast-Wave model fit as discussed in next section.

       \begin{table}[htbp]
       \caption{Fit parameters: slope temperature $T_{slope}$ from Boltzmann fitting and freeze out temperature $T_{f}$ from Blast-Wave fit. }
       \label{bzfit_par}
       \centering
       \begin{tabular}{p{40pt} p{70pt} p{70pt}}
          \hline
          \hline
          $Cent.$   &$T_{slope}$(MeV)           & $T_{f}$(MeV)          \\
          \hline
          0-10\%   &93.06$\pm$0.91        &61.60$\pm$0.81  \\
          10-20\%   &92.67$\pm$1.74        &62.99$\pm$0.98  \\
          20-40\%   &86.55$\pm$1.35        &64.64$\pm$0.89  \\
          40-60\%   &78.50$\pm$1.54        &64.51$\pm$0.76  \\
          60-80\%   &73.87$\pm$1.62        &60.58$\pm$1.05  \\
          \hline
          \hline
       \end{tabular}
       \end{table}

%%%%%%%%%%%%%%%%%%%%%%%%%%%%%%%%%%%%%%%%%%%%%%%%%%%%%%%%%%%%%%%%%%%%%%%%%%%%%%%%%%%%%%%%%%%%%%%%%%%%%%%%%%%%%%%%%%%%%%%%%%%%%%%%%%%%%%%%%
\subsection{The Blast-Wave fit}

Considering that the Boltzmann function cannot fit well the $m_{T}$ spectra in the low $m_{T}$ region due to the existence of collective radial flow, we adopt a hydrodynamically inspired ``Blast-wave" model to describe the $m_{T}$ spectra with two free parameters: collective transverse flow velocity $\beta$ and kinetic freeze-out temperature $T_{f}$ . In this model, the spectrum is computed by boosting the thermal source both in longitudinal and transverse direction~\cite{Siemens,BW_ESch}. The radial velocity distribution $\beta_{r}$ in the region $0\leq \beta_{r} \leq R$ is described by a self-similar profile, which is parameterized by the surface velocity $\beta_{S}$:
       \begin{equation}
       \beta_r = \beta_{S} \cdot(\frac{r}{R})^\alpha,
       \label{radialflow}
       \end{equation}
where R is the freeze-out radius, namely the maximum radius of the expanding source at thermal freeze-out time.  $\beta_{S}$ is the particles' radial velocity on the surface of the freeze out volume when r = R, and the exponent $\alpha$ represents for the radial flow profile which describes the evolution of the radial flow velocity (when $\alpha$ =0, it means uniform velocity; when $\alpha$ =1, the expansion is similar to Hubble's law; when $\alpha$  = 2, it corresponds to a hydromechanical expansion).
The particle spectra are the superposition of individual thermal source at different r, each boosted with the angle $\rho=tanh^{-1}\beta_{r}(r)$:
       \begin{equation}
       \frac{dn}{m_{T}dm_{T}}\propto \int_0^{R} rdr m_{T}I_{0}(\frac{p_{T}sinh\rho}{T_{f}})
       K_{1}(\frac{m_{T}cosh\rho}{T_{f}}),
       \label{eq11}
       \end{equation}
where $K_{1},I_{0}$ are the modified Bessel function, $T_{f}$ is the freeze-out temperature. The spectrum shape is determined by the freeze-out temperature $T_{f}$, the velocity of the transverse expansion $\beta_{S}$, the flow velocity profile $\alpha$ and the mass of the particle $m_{0}$. The average flow velocity is estimated by taking an average over the transverse geometry: $\left\langle\beta_{r}\right\rangle=\beta_{S}\frac{2}{2+\alpha}$.

The spectra of protons can be well described by  the Blast-Wave model as shown in  Fig.~\ref{pr_fit}(b) and the fit parameters are listed in Table~\ref{bwfit_par}. Obviously, the whole spectra are well  fitted, even for the ``shoulder structure'' at low $m_{T}$ in central collisions (for the centralities of 0-10\% and 10-20\%). From central collisions to peripheral ones, the velocity $\beta_{S}$ decreases and while the temperature has only a slight increase~\cite{BW_BIAb}. It is noted that the similar trend was also seen in the fitting result in RHIC energy region, which might be owing to non-equilibrium effect in peripheral collision. For example, by replacing the Boltzmann statistics with the Tsallis statistics in traditional Blast-Wave model, the Tsallis Blast-Wave model (TBW) can describe the peripheral spectra well and give a more reasonable fitting result (a low value of $\beta_{S}$ and $(q-1)>0$ at peripheral collisions)~\cite{BW_Tang}.

The main results from the fitting of the $m_{T}$ spectra are following:

1. Radial flow ($\langle\beta_r\rangle$).

In Fig.~\ref{pr_fit}(b), the proton spectra have been described very well with the Blast-Wave model. The average radial flow is about 0.33c in the central collisions (0-10\%), and reduces to 0.22c in peripheral collisions (60-80\%). This result with a larger radial flow existing in central collisions is consistent with the physics picture of HICs.

\begin{figure}
      \includegraphics[width=0.48\textwidth]{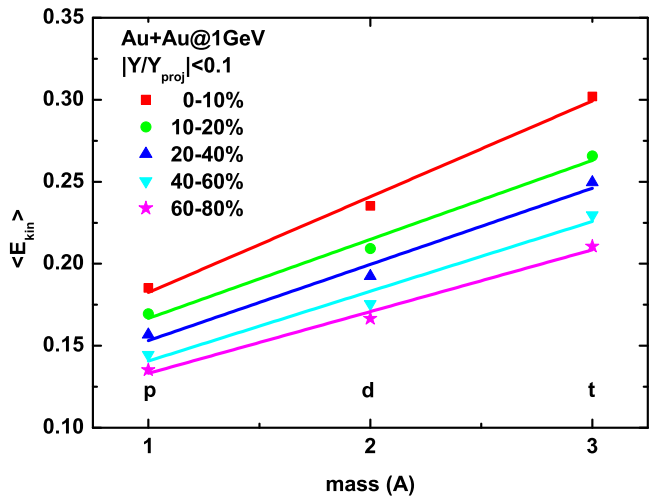}\\
      \caption{(Color online)  Average $E_{kin}$ of Z=1 particle in Au + Au at 1$A$ GeV  for different centralities (0-10\%,10-20\%,20-40\%,40-60\%,60-80\%), the solid line represents the linear fit on $\langle  E_{kin} \rangle$ vs A in each centrality, the fit parameters are displayed in Table~\ref{bwfit_par}}.
      \label{Ekin_A}
      \end{figure}

In another scenario, the average kinetic energy $\langle E_{kin}\rangle$ of protons, deuterons and tritons at mid-rapidity  are consistent with the picture described by the Blast-Wave model. Fig.~\ref{Ekin_A} shows the mass number dependence of $\langle E_{kin}\rangle$. They have a linear relationship: $\langle E_{kin}\rangle = E_{th}+ E_{flow} = \eta \cdot A+3T/2$ where $\eta = 1/\sqrt{1-\langle \beta^2_{r}\rangle}-1$. The slope is the quantity related to radial flow, and the intercept is the  quantity related to thermal temperature. We can also extract the radial flow and temperature information by the linear curve about $E_{kin}$ vs $A$, the result is listed in the lower part of the Table~\ref{bwfit_par}, which is comparable with the Blast-Wave fit result.

       \begin{table}[htbp]
       \caption{The parameters of radial flow $\beta_S$, $\langle\beta_r\rangle$ and the freeze-out temperature $T_{f}$ at various centralities with different fitting methods. The upper one is from proton $m_{T}$ spectra described by the Blast-Wave model. The lower one is from fitting the average kinetic energy $\langle E_{kin}\rangle$ of p, d, t, which is displayed in Fig.~\ref{Ekin_A}.}
       \label{bwfit_par}
       \centering
       \begin{tabular}{p{40pt} p{50pt} p{70pt} p{50pt}}
       \hline
       \hline
                    & $\beta_S$   &$\langle\beta\rangle$   & $T_{f}$ (MeV)  \\
       \hline
                    &Blast-wave                                \\
       \hline
       %20w event fit
            0-10\%  &0.3615       &0.3309$\pm$0.0028      &61.60$\pm$0.81  \\
           10-20\%  &0.3199       &0.2928$\pm$0.0037      &62.99$\pm$0.98  \\
           20-40\%  &0.2798       &0.2561$\pm$0.0039      &64.64$\pm$0.89  \\
           40-60\%  &0.2452       &0.2241$\pm$0.0039      &64.51$\pm$0.76  \\
           60-80\%  &0.2355       &0.2155$\pm$0.0056      &60.58$\pm$1.05  \\
       \hline
                    &$\langle E_{kin}\rangle$ vs A                                    \\
       \hline
            0-10\%  &             &0.3275$\pm$0.0047      &82.72$\pm$6.97  \\
           10-20\%  &             &0.2996$\pm$0.0057      &79.02$\pm$6.93  \\
           20-40\%  &             &0.2946$\pm$0.0070      &71.12$\pm$8.94  \\
           40-60\%  &             &0.2752$\pm$0.0032      &67.52$\pm$7.69  \\
           60-80\%  &             &0.2670$\pm$0.0036      &63.63$\pm$5.27  \\
       \hline
       \hline
       \end{tabular}
       \end{table}

2. Freeze out temperature ($T_{f}$)

We use Boltzmann distribution to fit the proton spectra, with the fitting parameters at different centralities called ``slope temperature"  listed in Table~\ref{bzfit_par}. The temperature shows a decreasing trend from 93.06 MeV in central collisions to 73.87 MeV in peripheral collisions. Comparing with the Boltzmann fit results, the freeze-out temperature from the Blast-Wave fit is smaller, and shows a little increasing trend. This difference is owing to that the collective radial flow motion effect has been misunderstood as the thermal part of motion in the Boltzmann description.
%It is worth noting that the $T_{f}$ increases slightly with the centrality, which is also observed in RHIC energy and it may be due to the non-equilibrium effect~\cite{BW_BIAb,BW_Tang}.\\

3. Flow profile ($\alpha$)

The $\alpha$ exponent describes the evolution of the radial flow velocity from any radius r ($0<r<R$). We set the parameter $\alpha$ free in Blast-Wave fit, which is different from some previous works keeping a fixed value equal to 1 or 2~\cite{BW_ESch,AGS_rflow3,BW_Tang} or some assuming a common flow (i.e. $\alpha=0$)~\cite{Lisa_rflow,flow_evi2,Hartnack_rflow4,FuF_rflow5}.
The $\alpha$ makes a little difference to the fitting value of $\beta_{S}$ while the choice of R value have no influence on the fitting which chosen R = 40 fm in our work. We get the $\alpha$ value equal to 0.336 on the spectra of 0-10\% centrality and fix this value when do the fitting on the other centralities. A similar value of $\alpha$ has been also obtained in ~Ref.~\cite{BW_Oana} for Au + Au collisions in ultra-relativistic energy.

%%%%%%%%%%%%%%%%%%%%%%%%%%%%%%%%%%%%%%%%%%%%%%%%%%%%%%%%%%%%%%%%%%%%%%%%%%%%%%%%%%%%%%%%%%%%%%%%%%%%%%%%%%%%%%%%%%%%%%%%%%%%%%%%%%%%%%%%%
%%%%%%%%%%%%%%%%%%%%%%%%%%%%%%%%%%%%%%%%%%%%%%%%%%%%%%%%%%%%%%%%%%%%%%%%%%%%%%%%%%%%%%%%%%%%%%%%%%%%%%%%%%%%%%%%%%%%%%%%%%%%%%%%%%%%%%%%%
\section{Nuclear modification factor}

To explore the nuclear medium response, the ratio $R_{CP}$ between the particle yield in central collisions and the particle yield in peripheral collisions, has been introduced. Both yields are normalized by corresponding nucleon-nucleon binary collision numbers $\langle N_{coll}\rangle$ (binary scaling):
       \begin{equation}
       %R_{AA}=\frac{1}{N_{coll}^{AA}}\frac{Yield(AA)}{Yield(pp)}. \\
       R_{CP}=\frac{Yield(central)/\langle N_{coll}^c\rangle}{Yield(peripheral)/\langle N_{coll}^p\rangle},
       \label{eq12}
       \end{equation}
where the yield is the differential invariant yield ($\frac{d^{2}N}{2\pi p_{T}dp_{T}dy}$). If the nucleus-nucleus collision is a mere superposition of $N_{coll}$ independent nucleon-nucleon collisions,  $R_{CP}$ would be unit with $p_{T}$. Thus any departures from $R_{CP}$=1 indicate nuclear medium effects or other dynamical effects.

%%%%%%%%%%%%%%%%%%%%%%%%%%%%%%%%%%%%%%%%%%%%%%%%%%%%%%%%%%%%%%%%%%%%%%%%%%%%%%%%%%%%%%%%%%%%%%%%%%%%%%%%%%%%%%%%%%%%%%%%%%%%%%%%%%%%%%%%%
\subsection{The normal NMF}

\begin{figure}
       % Requires \usepackage{graphicx}
       \includegraphics[width=0.48\textwidth]{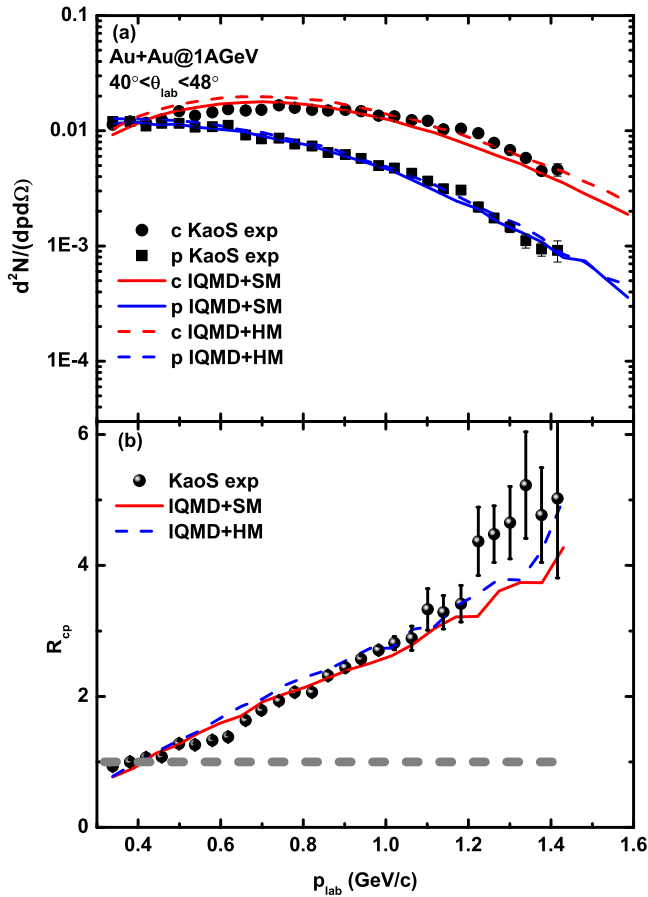}
       \caption{(Color online) a) The central and peripheral spectra of protons taken from the KaoS experimental data~\cite{CMuntz} are compared with our IQMD simulations with SM and HM interaction parameters. The character ``c" (``p") represents central (peripheral) collision, circle (square) represents the KaoS experimental result, solid red (blue) line represents IQMD with SM interaction and dash red (blue) line IQMD with HM interaction.  Central (circle symbol) and peripheral (square symbol) collisions are selected by the charge particle multiplicity, namely $M_{z}>50$ and $M_{z}\leq26$.
       b) Experimental $R_{cp}$ (circle) calculating from the KaoS experimental data is compared with IQMD with SM (solid red line) and HM (dash blue line) results. The dash line ($R_{CP}$=1) is just guiding the eyes.}
       \label{exp_nmf}
       \end{figure}

The normal $R_{CP}$ versus $p_{T}$ is obtained by dividing the $p_{T}$ spectra in central collisions with the ones in peripheral collisions with taking the respective binary collision number into account. To test the reliability of model, the results from IQMD calculation are compared with that extracted from KaoS experimental data. Following the experimental conditions ~\cite{CMuntz}, the central collision events are chosen with charged particle multiplicity $M_{z} > 50$, and the peripheral collision events with $M_{z} < 26$. The identified protons are chosen the same as the KaoS experiment, namely, in the polar angle of 40-48 degrees, and the momentum range 0.320-1.440 GeV/$c$. The spectra of proton in central and peripheral collisions are showed in Fig.~\ref{exp_nmf}(a), together with the result from IQMD+SM(HM) representing with different type lines elaborated in figure caption. Results demonstrate that both the soft and hard EOS can describe the experimental data nicely. Shown in Fig.~\ref{exp_nmf}(b) is the $R_{CP}$  as a function of total laboratory momentum from IQMD simulation and the KaoS experiment. Two parameter sets of equation of state, namely SM and HM, are employed in IQMD calculation. The $R_{CP}$ of protons from experiment (circle) and from IQMD calculation (solid red line: SM; dash blue line: HM) increase almost linearly with the total momentum in laboratory system ($p_{lab}$). This is consistent with those behaviors from lower RHIC Beam-Energy-Scan energies, e.g. the  $R_{CP}$ for proton in Au+Au at 7.7$A$ GeV \cite{Horvat_BES}, and can be understood by radial boosts and/or the Cronin Effect ~\cite{CroninE1}. The $R_{CP}$ of proton is not so sensitive to the hard or soft equation of state in the IQMD calculations,  indicating the multiple collision process dominates here. It's a  need to note that the Rcp from model are arbitrarily scaled to match with the experimental  result owing to the unknown $N_{coll}$ in KaoS experiment.

In IQMD model, two particles collide if the minimum relative distance of their centroids of the Gaussian wave during their motion, in the CM frame fulfills the requirement:
       \begin{equation}
       d \leq d_{0} = \sqrt{\frac{\sigma_{tot}}{\pi}},
       \label{eq13}
       \end{equation}
where the cross section $\sigma_{tot}$ is assumed to be the free cross section of the regarded collision type (N-N, $N-\Delta$, etc.).
In this work, the collision numbers in every centralities are counted with the ``$N_{collt}$" counter when each collision occurs. For each collision the phase space densities in the final states are checked in order to assure that the final distribution in phase space is in agreement with the Pauli principle, the Pauli blocking number has been counted by the ``$N_{paubl}$'' counter. Then, we can get the  real collision number ($N_{coll} = N_{collt}-N_{paubl}$). For  Au+Au collision at incident energy 1$A$ GeV, the average collision numbers are 242.5, 473, 795.5, 1136, 1463 in five centralities from  peripheral to central collisions in our IQMD model, respectively.
%Those numbers seems larger than the Glauber model estimation at 1$A$ GeV. However,  considering that the %Glauber model  ignores the collisions in spectators and while the IQMD transport model overestimates it, the larger %collision number $N_{coll}$ in IQMD is understandable. Also considering the self-consistency, the values of  %$N_{coll}$ from IQMD have been used.

\begin{figure}
       \includegraphics[width=0.50\textwidth]{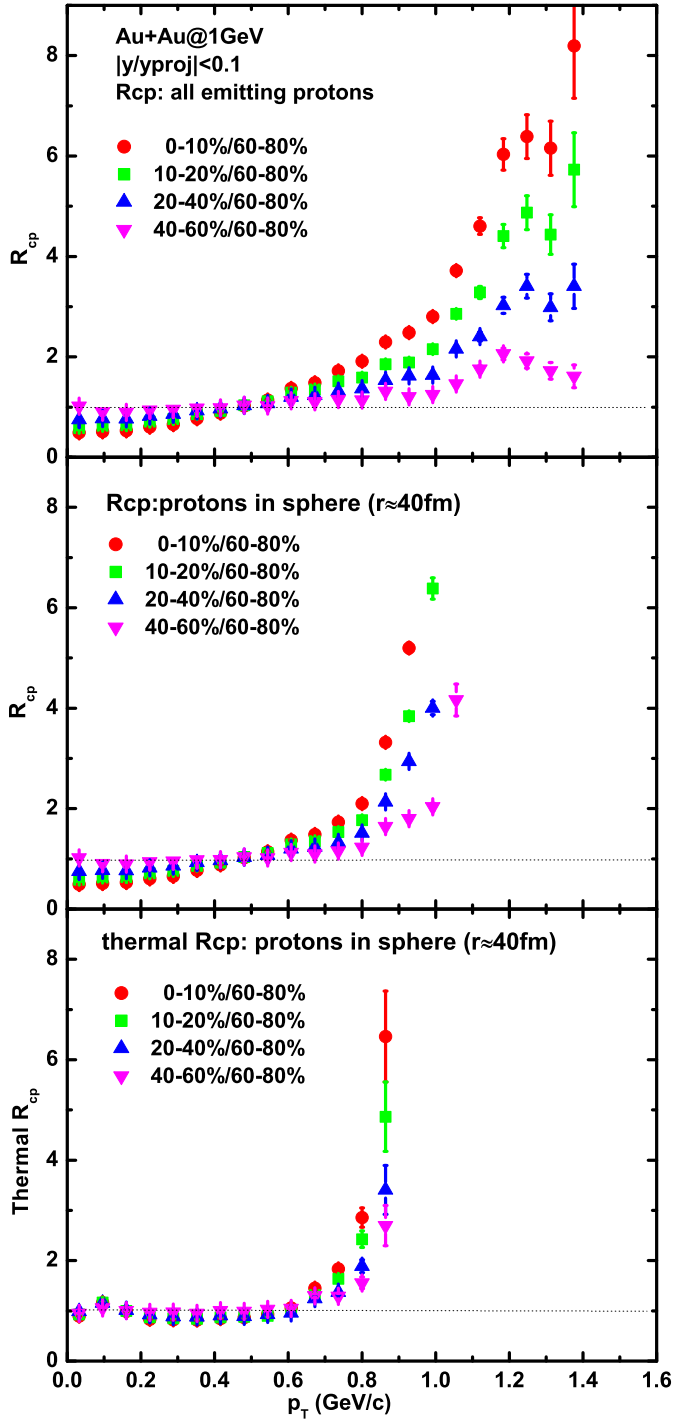}
       \caption{(Color online)  Nuclear modification factor, $R_{CP}$ of protons in Au + Au at 1$A$ GeV. The  panel (a) is the original $R_{CP}$, the panel (b) is the $R_{CP}$ inside the freeze-out sphere, and the panel (c) shows the $R_{CP}$ (b) after deducting the radial flow effect from  the panel (b), which is named as the thermal $R_{CP}$. The dot line ($R_{CP}$=1) is just guiding the eyes.}
       \label{pt_nmf}
       \end{figure}

The $R_{CP}$ versus $p_{T}$ shows centrality dependence. In Fig.~\ref{pt_nmf}(a), four cases at different centralities are investigated, which are $\frac{0-10\%}{60-80\%}$ (red dot), $\frac{10-20\%}{60-80\%}$ (green square), $\frac{20-40\%}{60-80\%}$ (blue up triangle) and $\frac{40-60\%}{60-80\%}$ (pink downtriangle). From the most peripheral case in numerator (i.e. $\frac{40-60\%}{60-80\%}$) to the most central case in numerator (i.e. $\frac{0-10\%}{60-80\%}$), the $R_{CP}$ becomes more and more $p_{T}$ dependent. For all these cases, $R_{CP}$ rises with $p_{T}$, with a cross point shows up at $p_{T}=$0.5 GeV/c, which may suggest a balance between two mechanisms for the $p_{T}$ dependence of $R_{CP}$.

On the one hand, in the low $p_{T}$ region ($p_{T}$ less than 0.5 GeV/c), radial flow takes a major role in central collisions, which pushes protons to higher $p_{T}$ region and results in the smaller $R_{CP}$ at low $p_{T}$. On the other hand, in the high $p_{T}$ region ($p_{T}$ greater than 0.5 GeV/c), the Cronin effect (nucleon multiple scattering effect) ~\cite{CroninE2,CroninE3} tends to transform the longitudinal momentum into the transverse momentum and increase the $p_{T}$ in central HIC, leading to the larger $R_{CP}$ at high $p_{T}$.

%%%%%%%%%%%%%%%%%%%%%%%%%%%%%%%%%%%%%%%%%%%%%%%%%%%%%%%%%%%%%%%%%%%%%%%%%%%%%%%%%%%%%%%%%%%%%%%%%%%%%%%%%%%%%%%%%%%%%%%%%%%%%%%%%%%%%%%%%
\subsection{The thermal NMF}

In order to investigate the thermal property of the medium in the collision overlapping region, one needs to deduct the contribution of collective flow to focus on  thermal effect. The radial flow parameter extracted by the Blast-Wave fit has been displayed in Table~\ref{bwfit_par}. By subtracting the radial velocity of each particle, we recalculate the $R_{CP}$ of protons without the radial flow contribution.

To determine the magnitude of radial flow, one needs to figure out the size of freeze-out volume. In the transport model IQMD, a density criterion is applied here. During the expansion process, the stage can be chosen as freeze-out stage, when the density at the center of each collision reaches $1/8\rho_{0}$. At this stage, all the products  (including fragments and free nucleons) cease the strong interaction among them and almost fix their momenta. For the case of Au+Au collision at $E_{beam}$ = 1$A$ GeV, this freeze-out time is about 60 fm/c, and the corresponding radius $R_{max}$ of freeze-out volume is about 40 fm (half of the reducing edge of radius distribution of all emitting protons), which is of course model dependent.  Fig.~\ref{pt_nmf}(b) shows the NMF of protons inside the freeze-out volume.  In comparison with  Fig.~\ref{pt_nmf}(a), freeze-out sphere cuts off most of higher $p_T$ protons. Once the freeze-out sphere is fixed, the radial flow $\beta_{r}$ for the particles inside the volume can be calculated through formula in Eq. (\ref{radialflow}), here $\alpha$ = 0.336 is taken from the Blast-Wave fit for the $p_T$ spectra.
The contribution of radial flow on $p_{T}$ is the projection of radial momentum ($p_{r} = m_{0}\cdot \beta_{r}\cdot \gamma_{r}$) on the $p_{T}$ vector, and the thermal transverse momentum is $p_{T} = p_{T}-\frac{\vec{p_{r}}*\vec{p_{T}}}{|\vec{p_{T}}|}$. The thermal $R_{CP}$, is then obtained by dividing the thermal $p_{T}$ distribution in central collisions to that in the peripheral collisions with taking the respective binary collision number into account, which is shown in the Fig.~\ref{pt_nmf}(c). It is found that the thermal $R_{CP}$ becomes almost unitary below 0.6 GeV/c,  which illustrates that the original increasing $R_{CP}$ behavior in low $p_T$ region essentially stems from the collective radial flow effect, and the thermal motion of nucleus-nucleus collision can be seen as the independent overlap of nucleon-nucleon collisions. It is also noted that the thermal $R_{CP}$ is larger than 1 at $p_T$ above 0.6 GeV/c, where the Cronin effect plays a dominant role for the $R_{CP}$ increasing behavior versus $p_T$  ~\cite{CroninE1,CroninE2}.

%%%%%%%%%%%%%%%%%%%%%%%%%%%%%%%%%%%%%%%%%%%%%%%%%%%%%%%%%%%%%%%%%%%%%%%%%%%%%%%%%%%%%%%%%%%%%%%%%%%%%%%%%%%%%%%%%%%%%%%%%%%%%%%%%%%%%%%%%
%%%%%%%%%%%%%%%%%%%%%%%%%%%%%%%%%%%%%%%%%%%%%%%%%%%%%%%%%%%%%%%%%%%%%%%%%%%%%%%%%%%%%%%%%%%%%%%%%%%%%%%%%%%%%%%%%%%%%%%%%%%%%%%%%%%%%%%%%
\section{Conclusion}

In summary, the nuclear modification factor has been introduced in intermediate energy HIC in the framework of transport model, and thermal NMF has been proposed.
First, we test the reliability of the IQMD model by comparing the kinetic energy spectra of light fragments (p, d, t) obtained by the IQMD model with the EOS experimental data, and we found the model can describe the data nicely. Secondly, the transverse mass spectra of protons are fitted by the Boltzmann distribution function as well as the distribution function from Blast-Wave model. It is found that the latter gives a satisfying description of the transverse mass $m_{T}$ spectra of protons at freeze-out time, which demonstrates that the kinetic energy of protons contains the collective radial motion part and random thermal motion part. The fitting radial flow parameter $\beta_S$ and temperature parameter $T_{f}$ are consistent with the results from the EOS collaboration~\cite{Lisa_rflow} and the FOPI collaboration~\cite{FOPI_rflow1,FOPI_rflow2}. While, the Boltzmann description overestimates the proton yield at low $m_{T}$, because of the missing radial flow contribution. Thus the fit temperature from the Boltzmann description can be interpreted as the apparent temperature of the emission source, which decreases from the central collisions to the peripheral collisions.
It is found that both the radial flow effect and Cronin effect play their corresponding roles in shaping the $p_T$ dependence of $R_{CP}$. The radial flow magnitude can be extracted by fitting the $m_T$ spectra with the distribution function from Blast-Wave model. The thermal modification factor is then obtained by removing the contribution from the radial flow. It is found that the thermal $R_{CP}$ of protons is close to 1 at lower $p_{T}$, where protons emitted from Au+Au collisions  can be regarded as  independent superposition of emitted protons by nucleon-nucleon collisions, and while  $R_{CP}$  enhances at higher $p_{T}$, where the Cronin effect, i.e. multiple production mechanism of protons plays an increasing role. On the other hand, in light of this study,  the nuclear modification factor at very high $p_T$ which has been often used to investigate jet-quenching effect at RHIC and LHC energies was actually nearly not affected by the collective radial flow.

\section*{Acknowledgements}
This work was supported in part by the Major State Basic Research Development Program in China
under Contract No. 2014CB845401, the National Natural Science Foundation of China under Contract Nos.
11035009 and 11220101005,  the Knowledge In-
novation Project of Chinese Academy of Sciences under
Grant No. KJCX2-EW-N01.

\end{document}